\documentclass[10pt,letterpaper]{article}
\usepackage{opex3}
\usepackage{amsmath}
\usepackage{bm}
\DeclareGraphicsExtensions{.eps}

\begin{document}

%% NOTE: TITLE PAGE & TOC NOT USED FOR MANUSCRIPT SUBMISSIONS %%
%\title{Twin-Photon Confocal Microscopy}

%\vskip4pc

%\tableofcontents
%\clearpage
%% NO TITLE PAGE FOR OPEX SUBMISSIONS %%

%% START HERE
%%%%%%%%%%%%%%%%%% title page information %%%%%%%%%%%%%%%%%%
\title{Turbulence mitigation in phase-conjugated two-photon imaging}

\author{D.S. Simon$^{1,2,\ast}$ and A.V. Sergienko$^{1,2,3}$}

\address{$^1$ Dept. of Electrical and Computer Engineering, Boston
University, 8 Saint Mary's St., Boston, MA 02215.

$^2$ Photonics Center, Boston
University, 8 Saint Mary's St., Boston, MA 02215.

$^3$ Dept. of Physics, Boston University, 590 Commonwealth Ave.,
Boston, MA 02215.}

\email{$^\ast$ simond@bu.edu} %% email address is required



\begin{abstract}
It is shown that the use of phase conjugation in one arm of a correlated two-photon imaging apparatus allows
undistorted ghost imaging through a region with randomly-varying phase shifts. The images are formed from correlated pairs of photons
in such a way that turbulence-induced phase shifts gained by the photons during passage through the medium
cancel pairwise.
\end{abstract}


\section{Introduction}\label{intro}

In correlated two-photon imaging \cite{pittman}, also known as ghost imaging, images are constructed by means of spatial correlations between
pairs of photons. These pairs may be quantum-mechanically entangled photons
produced via parametric downconversion, or spatially correlated pairs from a classical light source
\cite{bennink1,shih,bennink2,gatti,scarcelli1,cai,ferri,valencia,zhang2,scarcelli2}.
In each pair, one photon (the target photon) interacts with an object, then strikes a single-pixel detector with no spatial resolution. The other photon (the reference photon) propagates freely to a CCD camera or other form of spatially resolving detector, without ever encountering the object.
Although neither photon is capable of producing an image of the object by itself, the image may be reconstructed from the spatial correlations between them when the pairs are detected in coincidence.

Recently, a number of theoretical and experimental investigations have looked at how ghost imaging is affected by turbulence in the propagation paths \cite{cheng,zhang,li,meyers1,meyers2,dixon}. In this paper, we alter the standard ghost imaging configuration by the addition of a phase conjugate mirror in one path and by partially merging the target and reference branches of the apparatus so that both photons experience the {\it same} turbulent conditions. We show that this alteration eliminates the effect of the randomly varying turbulence-induced phase shifts on the image.

The basic strategy proposed here, which consists of combining coincidence detection of correlated photon pairs with phase conjugation, is more general than the specific application (transmissive ghost imaging) given in this paper. It may be used for other types applications involving transmission of temporally- or spatially-modulated signals across a distance through a turbulent medium. The problem of optimizing optical communication through the turbulent atmosphere is a longstanding problem \cite{shap1,shap2,shap3}; a variation of the method discussed here which is appropriate for undistorted signal transmission from one side of a turbulent medium to the other is possible and will be discussed elsewhere.

{\bf Phase conjugation.} A phase conjugate mirror (PCM) \cite{fisher,zel1,pepper1}
is a nonlinear optical device for reversing the phase of a propagating light wave. More specifically, if the incoming complex electric field of the wave is $E({\mathbf x})e^{-i\omega t}$, the outgoing field after reflection is complex conjugated except for the time dependence, which is unchanged: $E^\ast ({\mathbf x})e^{-i\omega t}$. Phase conjugate mirrors may be constructed to operate via either stimulated Brillouin scattering or four wave mixing over some range of frequencies determined by a set of phase-matching conditions. One reason why PCMs are useful is that they exhibit a well-known cancelation of phase distortions. This effect can be described as follows. Suppose that a set of incoming wavefronts is distorted during passage through some region; for example, they may experience aberration while passing through an optical imaging system or there may be variations in the refractive index of the propagation medium. These distortions may be viewed as the result of spatially varying phase shifts added to the field.  After receiving these phase distortions, suppose that the wave reflects off a PCM and passes through the distorting region a second time. An identical set of phase distortions occur on the return trip, canceling the complex-conjugated phase distortions from the first passage. This phase cancelation effect has a number of applications
and has been used in the past to mitigate the effects of turbulence in imaging and signalling systems, with a number of different methods developed \cite{yariv, fischer, feinberg, ikeda1,ikeda2,pepper2, macdonald, alley,zhang1, kramer, statman, yau}
using four-wave mixing or dynamic holography.
Each method has its advantages compared to the others, but each also has drawbacks. For example, some of the proposed methods require the signal to make a round-trip back to its starting point, or require active cooperation between sender and receiver. Some methods work only for distorting media that are thin enough to completely image into a phase-conjugate mirror, while others work only for static aberrations or only for aberrations changing with very short characteristic time scales. Still others require a second reference beam that must remain coherent with the signal beam.

{\bf A new approach.} The method proposed here combines phase conjugation with a ghost imaging approach in a manner that eliminates the drawbacks mentioned above. For example, the photons are detected pairwise (similar to previous schemes involving reference beams), but the photons in the detected pairs are automatically coherent with each other due to their correlated production and the use of coincidence detection. Similarly, beyond the initial setup of appropriate sources and detectors at the two ends, there is no need for active cooperation between the ends. No round trip is needed, so information may be transmitted from one side of the turbulent medium to the other. Rather than sending a single photon across the medium and back, the idea here is to send two correlated photons through the medium just once, with no return trip, arranging for distortions to cancel between the two of them.

The scheme described here is part of a progression of methods (dispersion cancelation \cite{franson, steinberg, minaeva1},
aberration cancelation \cite{bonato1, bonato2, simon1},
etc.) that have been developed using classically correlated light beams or quantum mechanically entangled photon pairs to cancel various types of optical distortions in a variety of different situations. Since turbulence may be viewed as a form of aberration that varies randomly in time, the work here is a logical next step in this progression. In particular, if we remove the phase conjugate mirror from the apparatus described below (fig. \ref{pcmfig}), the resulting device is
essentially the same as that used in \cite{simon2} to cancel odd-order aberrations induced by an optical imaging system.
The passage of the photons through the turbulent medium while preserving the total phase of the biphoton wavefunction can be viewed as the result of having performed the measurement on a decoherence-free subspace \cite{lidar,walton2} of the two-photon system.

This approach bears similarities to ideas that have appeared in other contexts. For example, the plug and play quantum cryptography system of \cite{muller}, which required a single photon to undergo a round trip, was altered in \cite{walton} to allow a pair
of entangled photons to accomplish the same goal in a single one-way trip, making use of the stable phase relation between the entangled photons.
Similarly, Erkmen and Shapiro \cite{erkmen,gouet} used phase conjugation to cancel the phase dependence of photon pairs in order to simulate
dispersion-canceled quantum optical coherence tomography (QOCT) with classical states of light and second-order interference.

{\bf Outline of the paper.} We proceed as follows. In section \ref{background}, we briefly review ghost imaging and quickly survey the work done to date on ghost imaging through turbulence. Then we describe two types of lensless phase-conjugated ghost imaging: with the two photons experiencing independent turbulent conditions in section \ref{turbindep} and with both feeling the same turbulent conditions in section \ref{turbsame}. The desired turbulence cancelation will appear in the latter case.
In the following, we will assume for the sake of specificity that the illumination is provided by a downconversion light source. However, the entanglement of the downconverted pairs will play no role, so there appears to be no physical obstacle to using spatially-correlated classical light beams or a pseudothermal speckle source instead, as will be discussed in section \ref{classicalsection}. Conclusions and a brief mention of one potential application follow in section \ref{conc}.

\section{Ghost imaging and turbulence}\label{background}

{\bf Ghost imaging.}
In 1995, it was demonstrated experimentally \cite{strekalov} (based on theoretical work in \cite{klyshko, belinskii}) that if a double slit was placed in one of a pair of beams originating from downconversion, no interference pattern would be formed in that beam (due to its insufficient coherence), but that the interference effects would reappear if the coincidence detection rate {\it between} the two beams was measured; coherence is maintained for the pair of beams as a whole. This effect became known as {\it ghost interference} or {\it ghost diffraction}, with the word "ghost" referring to the seemingly spooky nonlocal nature of the effect.
The first demonstration of the related effect of {\it ghost imaging} was made soon after in \cite{pittman}, using frequency-entangled photon pairs generated by type II downconversion.
%A schematic diagram of the apparatus is shown in fig. \ref{stanghostfig}.

%\begin{figure}
%\centering
%\includegraphics[totalheight=2in]{standardghost}
%\caption[Ghost imaging with entangled photons]{\textit{Ghost imaging with entangled photon pairs. }}\label{stanghostfig}
%\end{figure}

In ghost imaging, a light source produces entangled photon pairs or spatially correlated pairs of light beams. One member of each pair (the target photon or target beam) is transmitted through an object, then detected by bucket detector $D_2$. $D_2$ should be large enough to collect all of the signal photons arriving at the far end of the apparatus. The detector registers whether photons passed through the object or were blocked; but since it has no spatial resolution an image can not be reconstructed by using the information from this detector alone.

The other member of the pair (the reference photon or reference beam) travels unobstructed to $D_1$, a detector capable of spatial resolution. There are variations of the setup with or without lenses in the reference branch, with the distances in the setup satisfying an appropriate imaging condition in each case.
Although $D_1$ allows spatial structure to be recorded, the photons reaching it have not interacted with the object, so that once again information from $D_1$ alone will not be sufficient to reconstruct the image.
However, when the information from the two detectors is combined via coincidence counting, the image reappears as the coincidence rate is plotted versus position in $D_1$.

In \cite{bennink2}, it was shown that ghost imaging can be carried out with classically correlated beams in place of entangled photon pairs. This experiment was the first indication that the essential element in ghost imaging is the spatial momentum correlation of the photons, not the entanglement.
In refs. \cite{gatti,cai}, the question was raised as to whether ghost imaging could be carried out with partially coherent thermal light from a classical source. This was successfully done in the experiments of refs. \cite{ferri,valencia}.
Other variations on ghost imaging that have appeared recently include {\it computational ghost imaging}
\cite{shapirocomp,bromberg} and {\it compressive ghost imaging} \cite{katz}.

\begin{figure}
\centering
\includegraphics[totalheight=2.0in]{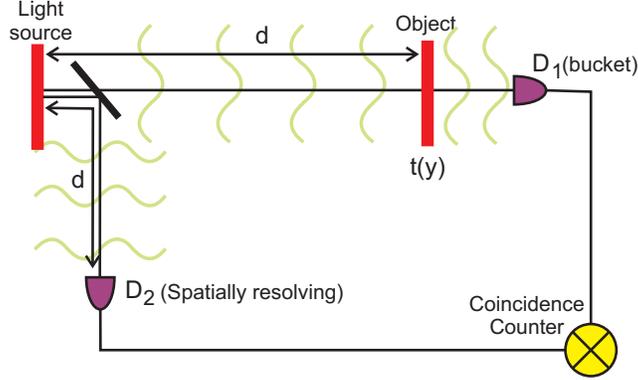}
\caption{\textit{Lensless ghost imaging setup used in \cite{cheng,li,zhang}.
Turbulence may appear in all portions of the optical paths of each beam. }}
\label{chengfig}\end{figure}

{\bf Ghost imaging through turbulence.}
Several theoretical analyses have recently been conducted of ghost imaging in the presence of turbulence. Ref.
\cite{cheng} looked at the lensless ghost imaging apparatus shown in fig. \ref{chengfig}. A fully incoherent light source was assumed, with turbulence filling both propagation paths. Analytic expressions were obtained for the output of the system, and detailed numerical simulations were performed for the case where the turbulence exists only in the target path. In \cite{li}, a similar analysis was performed for a partially coherent source, while in \cite{zhang} the case where both detectors are spatially resolving was examined. Note for comparison with section \ref{turbsame}, that in fig. \ref{chengfig} the two branches are separated, so that the turbulent fluctuations in the branches are independent of each other.

On the experimental side, ghost imaging with and without lenses through thin layers of turbulence have been carried out
\cite{meyers1,meyers2,dixon}.
It has been found that in the case with lenses the effect of the turbulence is strongly dependent on the location of the turbulent layer, with the effect becoming more prominent as the turbulence is moved closer to the lens. In this paper, we will restrict ourselves to the case of lensless ghost imaging, which is slightly simpler to treat theoretically. Throughout this work, all beamsplitters are assumed to be 50-50, i.e. to have equal reflection and transmission coefficients.

In passing, it might be pointed out that in many cases turbulence between the object and detector $D_1$ in fig. \ref{chengfig} should have no effect on the ghost image . Since $D_1$ is a bucket detector that only registers the arrival of light, not its spatial distribution, any additional spatial scrambling of phases and wavefronts after the object becomes irrelevant. This statement will fail to be true, however, if the scintillation effects are sufficiently strong. In addition, if the distance to $D_1$ is much larger than the size of the $D_1$, then some of the photons will begin to miss the detector and the coincidence rate will drop; this fact may limit the distance allowed between the object and the bucket detector.

\section{Phase-conjugated ghost imaging with independent turbulent regions}\label{turbindep}

\begin{figure}
\centering
\includegraphics[totalheight=2.0in]{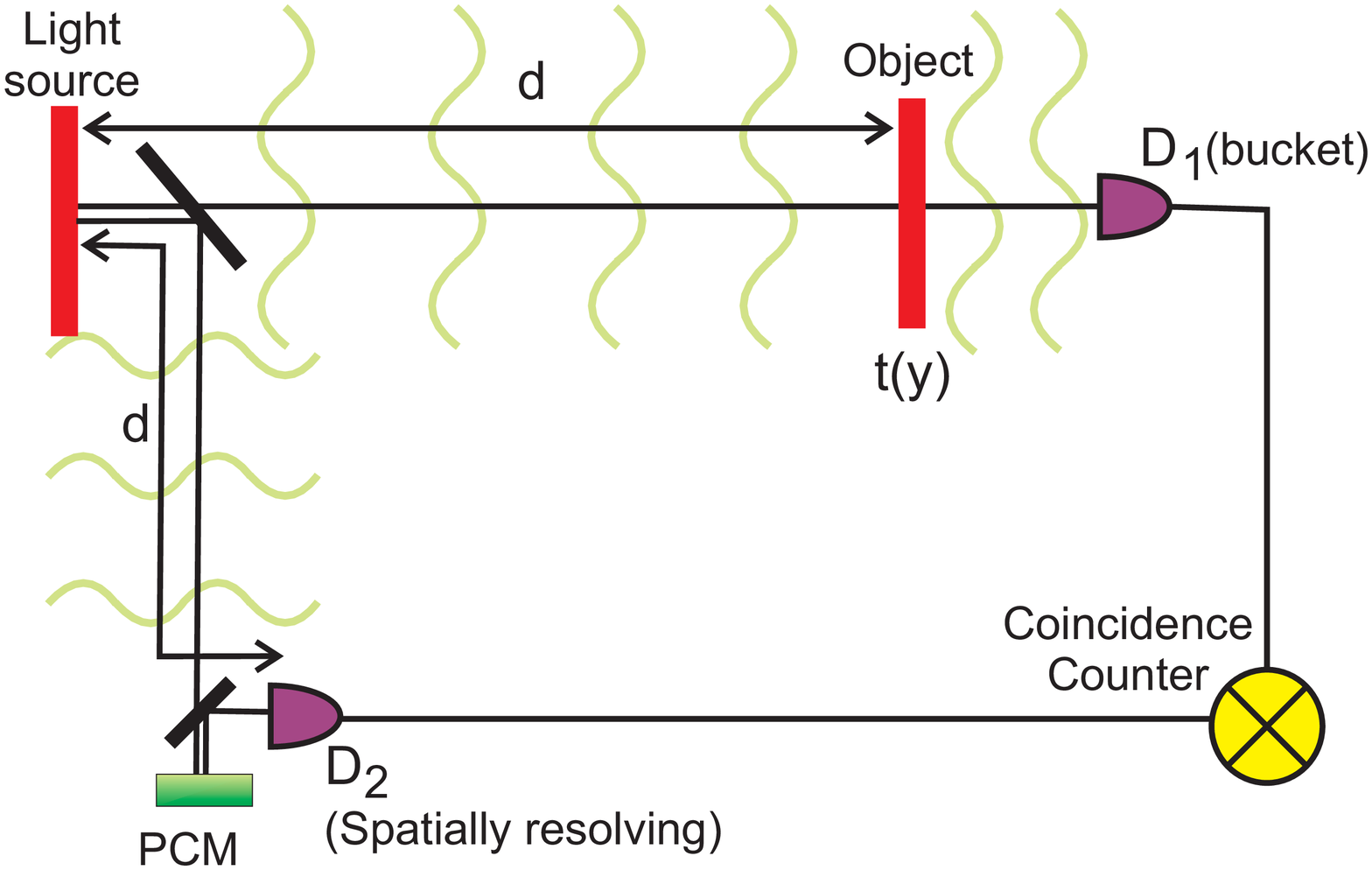}
\caption{\textit{Lensless ghost imaging setup of fig. \ref{chengfig}, with phase conjugate mirror added before one detector. It is assumed that the distance from the PCM to the detector is very small compared to $d$ and that there is negligible turbulence after the PCM. The total distances in the two arms are still equal.}}
\label{pcm0fig}\end{figure}

As a first step toward our goal, consider the apparatus shown in fig. \ref{pcm0fig}. This differs from that of fig. \ref{chengfig} by the addition of a phase conjugate mirror before the detector $D_2$. We assume that a turbulent medium fills the full propagation region and that the distance from the light source to the first beam splitter is negligible compared to $d$.

The free-space Huygens-Fresnel propagation function from point $\bm\xi$ in the source plane to point ${\mathbf x }$ in a plane a distance $z$ away is given by
\begin{equation}h_0(\bm\xi ,{\mathbf x })={1\over {i\lambda z}} e^{{{ik}\over {2z}}|\bm\xi -{\mathbf x}|^2}.\end{equation}
Here, we have neglected an overall constant (independent of
${\mathbf x}$ and $\bm\xi$) phase.
In the presence of turbulence, an extra factor $e^{\eta (\bm\xi ,x)+i \phi (\bm\xi ,x)}$ is introduced, \begin{equation}h_\phi (\bm\xi ,{\mathbf x })={1\over {i\lambda z}}e^{{{ik}\over {2z}}|\bm\xi -{\mathbf x})|^2}e^{\eta (\bm\xi ,{\mathbf x})+i\phi (\bm\xi ,{\mathbf x})}.\label{turbprop}\end{equation} The functions $\eta (\bm\xi ,{\mathbf x})$ and $\phi (\bm\xi ,{\mathbf x})$ vary randomly
with time, necessitating a time or ensemble average, to be denoted by angular brackets $\langle \dots \rangle$.
The method to be proposed here cancels only the phase fluctuations $\phi (\bm\xi ,x)$. Since the phases are our main focus, we will for simplicity
ignore the random amplitude variation or scintillation term $\eta (\bm\xi ,x)=0$ through most of this paper; in the conclusion we will briefly discuss the effect of reintroducing a nonzero scintillation term.

Since the two photons pass through different turbulent regions, turbulence in each branch will be described by a separate phase function, $\phi_1$ or $\phi_2$. For analytical simplicity we use the standard quadratic approximation to the $5/3$-power law correlation function, with spatially uniform structure function $C_n^2$. Pairwise correlations of the turbulence-produced phases are given by \begin{equation}
\bigg \langle \left( e^{i\phi_j({\mathbf x},{\mathbf y})}\right) \cdot \left( e^{ i\phi_j ({\mathbf x^\prime} , {\mathbf y^\prime} )}\right)^\ast\bigg\rangle =
\langle e^{i\phi_j({\mathbf x},{\mathbf y})-i\phi_j ({\mathbf x^\prime} ,{\mathbf y^\prime} )}\rangle =e^{-\alpha_j \left[ |({\mathbf x}-{\mathbf x^\prime} )|^2+ |{\mathbf y}-{\mathbf y^\prime} |^2+({\mathbf x}-{\mathbf x^\prime} )\cdot ({\mathbf y}-{\mathbf y^\prime} )\right] },\label{phicor}\end{equation} for $j=1,2$. The degree of turbulence in path $j$ is described by the parameter $\alpha_j={1\over {2\rho_j^2}}$, where $\rho_j=(1.09C_{n,j}^2k^2z_j)^{-3/5}$ is the turbulence coherence length \cite{ricklin}. Further, we assume in this section that the fluctuations in the two paths are statistically independent, so that the factorization
\begin{equation}\langle e^{i\phi_1(\bm\xi_1,{\mathbf x_1})-i\phi_2 (\bm\xi_2 ,{\mathbf x_2})-i\phi_1 (\bm\xi^\prime_1,{\mathbf x_1})
+i\phi_2 (\bm\xi^\prime_2 ,{\mathbf x_2})}\rangle \label{phifactor} =
\langle e^{i\phi_1 (\bm\xi_1 ,{\mathbf x_1})-i\phi_1 (\bm\xi^\prime_1,{\mathbf x_1})}\rangle\cdot \langle e^{-i\phi_2^\ast (\bm\xi_2 ,{\mathbf x_2}) +i\phi_2 (\bm\xi^\prime_2 ,{\mathbf x_2})}\rangle \end{equation} may be made.

The biphoton wavefunction leaving the crystal as assumed to have a Gaussian spatial profile with radius $a_0$ and fixed transverse coherence width $r_c$, \begin{equation}\psi_0(\bm\xi_s,\bm\xi_i)=\langle 0|\hat E_s^{(+)}(\bm\xi_s )\hat E_i^{(+)}(\bm\xi_i )|\Psi\rangle
\; =\; I_0 e^{-{{|\bm\xi_s-\bm\xi_i|^2}\over
{2r_c^2}}}e^{-{{|\bm\xi_s|^2 +|\bm\xi_i|^2}\over {4a_0^2}}} \;=\;I_0e^{-{{|\bm\xi_s |^2+|\bm\xi_i|^2}\over
{r_0^2}}}e^{{{\bm\xi_s\cdot \bm\xi_i}\over {r_c^2}}}, \label{ecor}\end{equation} where \begin{equation}{1\over {r_0^2}}={1\over {4a_0^2}}+{1\over {2r_c^2}},\end{equation} $I_0$ is a constant, and $|\Psi>$ is the two-photon part of the outgoing state.

Use of Eq. (\ref{turbprop}) tells us that
the impulse response function along the path leading to $D_1$ is \begin{equation}h_1(\bm\xi_1 ,{\mathbf x_2})= C_1 e^{{{ik}\over {2d}}| {\mathbf x_1}-\bm\xi_1|^2}e^{i\phi_1(\bm\xi_1, {\mathbf x_1})}t({\mathbf x_1}),\label{h1}\end{equation} where $C_1={1\over {i\lambda d}}$.
The impulse response for the path leading to $D_2$ will be
of the same form, except without the $t({\mathbf x_1})$ factor, and with the whole expression complex conjugated:
\begin{eqnarray}h_2(\bm\xi_2 ,{\mathbf x_2})&=& \left[ C_1 e^{{{ik}\over {2d}}| {\mathbf x_2}-\bm\xi_2|^2}e^{i\phi_2(\bm\xi_2, x_2)}\right]^\ast\label{h2}\\ &=& C_1^\ast e^{-{{ik}\over {2d}}| {\mathbf x_2}-\bm\xi_2|^2}e^{-i\phi_2(\bm\xi_2, x_2)}
\nonumber.\end{eqnarray} Let $\eta_1$, $\eta_2$ be the quantum efficiencies of the two detectors, and let $A_2$ be the (small) area of one detection cell of the spatially-resolving detector.
The coincidence rate is then given by \begin{equation} {\cal R}_c({\mathbf x_2})= \eta_1\eta_2A_2
\int G({\mathbf x_1},{\mathbf x_2}) d^2x_1 ,\label{rdef}\end{equation} where the correlation function \begin{equation}G({\mathbf x_1},{\mathbf x_2})=\langle \left| \psi ({\mathbf x_1},{\mathbf x_2})\right|^2\rangle\label{gdef}\end{equation} is the mean square of the biphoton amplitude in the detection planes,
\begin{equation}\psi ({\mathbf x_1},{\mathbf x_2})=
\int \psi_0(\bm\xi_1 ,\bm\xi_2)h_1(\bm\xi_1 ,{\mathbf x_1})h_2(\bm\xi_2 ,{\mathbf x_2})d^2\bm\xi_1d^2\bm\xi_2 \label{psisimple}.
\end{equation}

Substituting the propagation functions (eqs. \ref{h1} and \ref{h2}) into eqs. \ref{gdef}-\ref{psisimple}, we find that
\begin{eqnarray} G({\mathbf x_1},{\mathbf x_2})&=&\eta_1\eta_2A_2\left| C_1\right|^4 \left|t({\mathbf x_1})\right|^2\int \psi_0(\bm\xi_1 ,\bm\xi_2^\prime )\psi^\ast_0(\bm\xi_2 ,\bm\xi_1^\prime )\label{gexpression}\\
& & \qquad \times \; \langle e^{i\phi_1 (\bm\xi_1 ,{\mathbf x_1})-i\phi_2 (\bm\xi_2 ,{\mathbf x_2})-i\phi_1 (\bm\xi_1^\prime,{\mathbf x_1})
+i\phi_2 (\bm\xi_2^\prime ,{\mathbf x_2})}\rangle \nonumber \\
& & \qquad \times \; e^{{{ik}\over {2d}} \left[ \bm\xi_1^2+\bm\xi_2^{\prime 2}-\bm\xi_1^{\prime 2}-\bm\xi_2^2-2
{\mathbf x_1} \cdot(\bm\xi_1 -\bm\xi_1^\prime )+2
{\mathbf x_2} \cdot(\bm\xi_2 -\bm\xi_2^\prime )\right] }d^2\bm\xi_1 d^2\bm\xi_1^\prime d^2\bm\xi_2 d^2\bm\xi_2^\prime .\nonumber
\end{eqnarray} Making use of eqs. \ref{phicor}-\ref{ecor}, this becomes
\begin{eqnarray} G({\mathbf x_1}, {\mathbf x_2})&=& \eta_1\eta_2A_2I_0\left| C_1\right|^4 \left|t({\mathbf x_1})\right|^2\int e^{-{{|\bm\xi_1 |^2+|\bm\xi_1 |^{\prime 2}+|\bm\xi_2 |^2+|\bm\xi_2 |^{\prime 2}}\over {r_0^2}}} \\
& & \times \;e^{{\bm\xi_1\cdot\bm\xi_2^\prime}+{\bm\xi_2\cdot\bm\xi_1^\prime}\over {r_c^2}}
e^{{{ik}\over {2d}}\left[ \bm\xi_1^2+\bm\xi_2^{\prime 2}-\bm\xi_1^{\prime 2}-\bm\xi_2^2-2
{\mathbf x_1} \cdot(\bm\xi_1 -\bm\xi_1^\prime )+2
{\mathbf x_2} \cdot(\bm\xi_2 -\bm\xi_2^\prime )\right]}
\nonumber \\ & &  \qquad\times\;
e^{-\alpha_1 |\bm\xi_1-\bm\xi_1^\prime |^2}e^{-\alpha_2 |\bm\xi_2-\bm\xi_2^\prime |^2}d^2\bm\xi_1 d^2\bm\xi_1^\prime d^2\bm\xi_2 d^2\bm\xi_2^\prime .\nonumber\end{eqnarray} The $\bm\xi,\bm\xi^\prime $ integrals are Gaussian, and so can be easily evaluated.
The coincidence rate then takes the form \begin{equation}{\cal R}_{pcm}({\boldmath x_2})=Ce^{-|{\bm x_2|^2}/2W_{pcm}^2}\int d^2x_1 |t(x_1)|^2 e^{-{{|\bm x_1-m\bm x_2|^2}/{2R_{pcm}^2}}} .\label{coincresult0}\end{equation}

In the absence of turbulence ($\alpha_1=\alpha_2=0$), the correlation factor $m$
(measuring the spatial correlation of the outgoing beams), resolution $R$, and field of view $W$ are given by:
\begin{eqnarray}
2m_{pcm}&=&-{{r_0^2\left(\Delta_0-2k^2/d^2\right)}\over {r_c^2\Delta_0}}\label{mpcm}\\
2R_{pcm}^2&=&{{d^2r_0^2}\over {2k^2}}\left( {{\left(\Delta_0-2k^2/d^2\right)^2+\left( 4k/dr_0^2\right)^2 }\over{\Delta_0}}\right)\label{rpcm}\\
2W^2_{pcm} &=& {{d^2r_0^2}\over {2k^2}} \left({{\Delta_0\left[ \left(\Delta_0-2k^2/d^2\right)^2+\left(4k/dr_0^2\right)^2 \right]}\over{\Delta_0^2-\left( r_0^4/4r_c^4\right)\left( \Delta_0-2k^2/d^2\right)^2}}
\right)\label{wpcm}
,\end{eqnarray}
where \begin{equation}\Delta_0={1\over {4a_0^4}}\left( 1+{{4a_0^2}\over {r_c^2}}\right) +{{k^2}\over {d^2}}.\label{delta0}\end{equation}
For comparison, a similar calculation without the PCM (fig. \ref{chengfig}) gives a coincidence rate of the form
\begin{eqnarray}
2m_{0}&=&r_0^2/r_c^2\\
2R_{0}^2&=&{{d^2r_0^2}\over {2k^2}}\Delta_0\label{r0}\\
2W_{0}^2&=&{{2a_0^2d^2}\over{k^2}}\left({{2r_c^2}\over{2r_c^2+r_0^2}}\right) \Delta_0 .\end{eqnarray}

%In the absence of turbulence, we see that the widths satisfy \begin{equation}R_0^2\ge R_{pcm}^2.\end{equation} Thus, in terms of resolution, the apparatus with the PCM (fig. \ref{pcm0fig}) will always perform as well or better than the version without (fig. \ref{chengfig}), with the improvement increasing as the size of the illumination beam increases. However, in both cases, the width scales like $R\sim \sqrt{\alpha}$ for large $\alpha$, indicating increased degradation of the image with increased turbulence. Furthermore, in practice the improvement in the phase-conjugated case will usually be very small in far-field conditions ($ka_0r_c/d<<1$, with $k=2\pi/\lambda$).

The coincidence rates can be calculated for nonzero turbulence as well, but the resulting expressions are too complicated to provide much enlightenment. We state the results in the appendix for the sake of completeness, but refrain from discussing them in detail. We simply note that for both cases (with PCM and without), the resolution degrades at roughly the same rate with increasing turbulence strength, the resolvable width $R$ growing  roughly as $\sqrt{\alpha}$ for large $\alpha$. So at this point there seems to be no obvious benefit to including the phase conjugate mirror. However, we show in the next section that by making a change in setup the PCM can lead to dramatic improvement in resolution in the presence of turbulence.

\section{Phase-conjugated ghost imaging with merged paths}\label{turbsame}

\begin{figure}
\centering
\includegraphics[totalheight=2.0in]{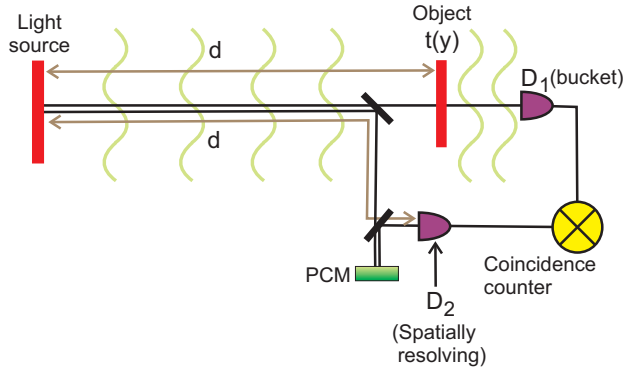}
\caption{\textit{Ghost imaging with a phase conjugate mirror and the two branches of fig. \ref{pcm0fig} merged, so that the two photons follow the same path through the turbulent region. It is assumed that the effects of any turbulence between the first beam splitter and the object are small enough to ignore. The same is assumed between the first beam splitter and $D_2$. The total distance from the light source to the object is the same as from the light source to $D_2$.}}
\label{pcmfig}\end{figure}

A crucial assumption in the previous section (as well as in refs. \cite{cheng,zhang,li}) is that the turbulent effects experienced by the two photons are statistically independent. This allowed the factorization of the four-fold expectation value in Eq. (\ref{phifactor}), which in turn allowed the evaluation of the integrals over the source
by means of the two-fold expectation in Eq. (\ref{phicor}).
We now remove the assumption of independence. This is accomplished in two steps. First, we move the beam splitter of fig. \ref{pcm0fig} from the source end of the turbulent region to the detector end as shown in fig. \ref{pcmfig}, so that both photons now move through the {\it same} turbulent region at the same time. Thus, we now have $\phi_1=\phi_2\equiv \phi $ and $\alpha_1=\alpha_2\equiv \alpha$.  Second, we assume that the two photons in each detected pair take very nearly the same path through the turbulent region; this can be accomplished for example by using light from {\it collinear} downconversion. In this case, the final detection
points for the two photons, ${\mathbf x_1}$ and ${\mathbf x_2}$, will be approximately equal, to within the distance allowed by diffraction. Similarly, in downconversion, the two photons are created at essentially the same location with the initial points $\bm\xi_1$ and $\bm\xi_2$ always within a submicron distance of each other. Thus, since the distance between the two initial points and between the two final points of the photons are both much smaller than the distance scale over which the turbulence-induced phase factors $e^{i\phi}$ vary significantly (typically on the order of centimeters to meters in the atmosphere, depending on the degree of turbulence and the propagation distance), we may take $\phi(\bm\xi_1,\bm x_1)\approx \phi(\bm\xi_2,\bm x_2)$.
The distances from the first beamsplitter to the object and the first beamsplitter to $D_2$ are assumed small compared to $d$, so that any turbulence in those regions will have little opportunity to affect the outcome.

For the case without the PCM, the coincidence rate will now be difficult to evaluate, since the above-mentioned factorization can no longer be done. However, for the PCM-based version of fig. \ref{pcmfig}, we find that all of the turbulent phase factors in Eq. (\ref{gexpression}) cancel.
Explicitly, the sum of phases that previously appeared in Eq. (\ref{gexpression}) now becomes \begin{eqnarray} & & i\phi (\bm\xi_1 ,{\mathbf x_1})-i\phi (\bm\xi_2 ,{\mathbf x_2})-i\phi (\bm\xi_1^\prime,{\mathbf x_1})+i\phi (\bm\xi_2^\prime ,{\mathbf x_2})\nonumber \\
& & \qquad \approx i\left[ \phi (\bm\xi_1 ,{\mathbf x_1})-\phi (\bm\xi_1 ,{\mathbf x_1})-\phi (\bm\xi_1^\prime,{\mathbf x_1})+\phi (\bm\xi_1^\prime ,{\mathbf x_1})\right]
\nonumber \\ & & \qquad =0,\label{phasecancel}\end{eqnarray} where we have again assumed that there are only phase (not amplitude) fluctuations. Thus, the random phases induced by the turbulence cancel exactly. The coincidence rate given in the appendix now reduces to the turbulence-free form, Eqs. (\ref{coincresult0}-\ref{wpcm}).

The reason for the cancelation effect is clear. As described in section \ref{intro}, it has long been known that
sending a wave front through a distorting region, reflecting it off a PCM, and sending it back through the distorting region produces an output (on the incident side) of a perfect time-reversed copy of the original, undistorted wavefront. The idea proposed here is similar, except that rather than sending a single wavefront through the distorting medium twice, the idea is to send two wavefronts through the medium once, then combine them (via coincidence counting) after inverting one in the PCM. Thus, no round-trip through the medium is necessary, making undistorted {\it one-way} transmission of images or other information possible. For full cancelation of the distorting effects, it is necessary that the two photons be affected equally by the medium; thus, the requirement that they follow the same path through the turbulent region. Further, to eliminate dispersive effects, narrow band spectral filters should be used before the detectors. Since turbulence does not create depolarization, either type I or type II downconversion will work; type II has the advantage that polarizing beam splitters may then be used to separate the signal and idler before detection.

\section{Classical versus quantum}\label{classicalsection}

Up to this point, we assumed that the illumination was provided by a parametric downconversion source, which may lead some readers to assume that the turbulence cancellation effect requires quantum entanglement. Entanglement is not, in fact, required. We wish to emphasize that the only essential ingredients for the turbulence cancellation are the following: (i) The detected photons must come in pairs that have gained equal phases during passage through the turbulent region. (ii) For these phases to be equal, it is necessary that they were emitted simultaneously from points close together in the source, and that they are detected at points that have nearly equal coordinates after emerging from the turbulent region. "Close together" in this context simply means separated by a distance much smaller than the characteristic size of the turbulent fluctuations. (iii) The unwanted phase of one of the photons should be reversed before detection.

Clearly, these requirements may be satisfied by classical means, with the paired photons replaced by paired locations in two classical beams. For example, a method similar to that of \cite{bennink1} may be employed for illumination: the
object could be scanned by a narrow beam, which is then split. One copy reaches the bucket detector after reflection or transmission from the object, while the
other copy is phase conjugated before reaching the spatially-resolving detector. As the scan progresses, an image will be built up, with the turbulent effects cancelling between the two copies of the classical beam.

Further, it should be noted that when a downconversion source is used for illumination there is no need for the outgoing photon pairs to be produced at a low enough rate for the pairs to be distinguishable from each other. Thus, we may apply strong pumping of the crystal, producing a high output flux of signal/idler pairs and replacing the coincidence detection by measurement of correlations between detector output currents as a function of position in detector 2.

{\bf PCM noise.} That we do not need to be in the low-brightness regime for the turbulence cancellation effect to occur is fortunate,
since a cursory look at the noise in the phase conjugate mirror shows that the illumination {\it must} be relatively bright.
If the annihilation operator of the input mode to the mirror is
$\hat a_{in}$, then the corresponding operator for the output is \begin{equation}\hat a_{out}=\sqrt{G-1}\hat a_{in} +\sqrt{G}\hat a_{noise},\end{equation}
where $G$ is the gain. Clearly, the average photon numbers in the input and output modes are related by \begin{equation}\langle \hat a_{out}^\dagger \hat a_{out}\rangle =(G-1)\langle \hat a_{in}^\dagger \hat a_{in}\rangle +G\hat \langle \hat a_{noise}^\dagger \hat a_{noise}\rangle
, \end{equation} or,
\begin{equation}N_{out}
= (G-1) N_{in} +GN_{noise}, \end{equation}
Typically it is assumed that the noise mode is in its vacuum state, so that $N_{noise}=1$. The treatment of the previous sections implicitly assumed that the input and output fields were roughly equal in magnitude, i.e. that $G\approx 2$. Thus, $N_{out}=N_{in}+2N_{noise}$. In order to prevent the noise from overpowering the signal, it is necessary that the number of photons in each input mode be large, $N_{in}>>1$. So if a downconversion source is used for illumination, the crystal must be strongly pumped, taking us away from the low-photon number regime where entanglement and other quantum effects would be visible.

\section{Conclusions}\label{conc}

{\bf Toward realistic applications.} We have shown that the use of phase conjugated ghost imaging has the potential to completely cancel distorting effects due to passage through a medium
with randomly varying phase shifts. Although we focused on the entangled-photon case of illumination by collinear parametric downconversion, the same mechanism will also work with a classical light source as long as sufficiently strong spatial correlations can be maintained between the two copies of the light. Thus, robust high-brightness classical sources should work, making the method more practical for real-world applications.

\begin{figure}
\centering
\includegraphics[totalheight=2.0in]{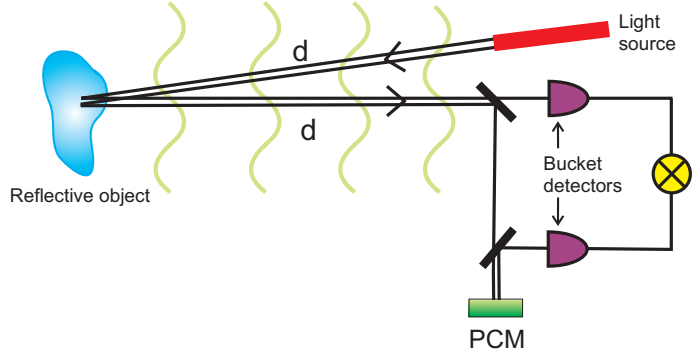}
\caption{\textit{Set-up for reflective stand-off sensing with cancellation of turbulent effects.}}
\label{reflectivefig}\end{figure}

One may envision a number of variations of the method designed to cancel turbulent effects in different situations. For example, a potentially useful application of ghost imaging is in reflective stand-off sensing of an object at a location remote from our apparatus \cite{meyers3,hardy}. The scheme shown in
fig. \ref{reflectivefig} displays one possible method for conducting remote sensing with turbulence cancellation. A narrow, highly-collimated beam of signal/idler pairs is used to scan the surface of a reflective object separated from our apparatus by a turbulent region. After reflecting back to the investigator's location, the beam is split and detected in coincidence after phase-conjugation of one copy. In this case, both beams are sent to bucket detectors; the spatial information about the target comes from scanning the narrow beam over the object. A more detailed discussion of this and other applications will be carried out elsewhere.

{\bf Scintillation.} One drawback of this method must be mentioned. Suppose we now restore the amplitude fluctuations, represented by the real factors $e^{\eta (\bm\xi ,x)}$ of Eq. (\ref{turbprop}). As mentioned in section \ref{turbindep}, we have up until now assumed that they were negligible. These terms introduce an extra exponential factor into Eq. (\ref{gexpression}) involving amplitude fluctuations. The analog of Eq. (\ref{phasecancel}) for these new terms is \begin{eqnarray} & & \eta (\bm\xi_1 ,{\mathbf x_1})+\eta^\ast (\bm\xi_2 ,{\mathbf x_2})+\eta^\ast (\bm\xi_1^\prime,{\mathbf x_1})+\eta (\bm\xi_2^\prime ,{\mathbf x_2})\nonumber \\
& & \qquad \approx\left[ \eta (\bm\xi_1 ,{\mathbf x_1})+\eta (\bm\xi_1 ,{\mathbf x_1})+\eta (\bm\xi_1^\prime,{\mathbf x_1})+\eta (\bm\xi_1^\prime ,{\mathbf x_1})\right]
\nonumber \\ & & \qquad =2\left[ \eta (\bm\xi_1 ,{\mathbf x_1})+\eta (\bm\xi_1^\prime ,{\mathbf x_1})\right].\label{amplitudenocancel}\end{eqnarray} Thus, rather than canceling, the scintillation terms are doubled in size, leading to increased twinkling. This method will therefore work best under conditions where the amplitude fluctuations are negligible and the turbulence can be treated as a fluctuating phase mask.

{\bf Computational imaging.} As a final observation, it may be noted that although computational and compressive ghost imaging are important areas of investigation with potential for a number of useful applications, the turbulence cancelation method described here will not work within the framework of a computational approach. Since both photons must pass through a turbulent region where the exact propagation functions are unknown, it is not possible to replace either photon by a simulation. So, the traditional two-detector version of ghost imaging clearly still has potential for results that can not be accomplished computationally, and it will continue to complement the computational approach into the future.

\section*{Appendix}

In section \ref{turbindep}, we outlined the method for calculating the coincidence rate for the situations shown in figs. \ref{chengfig} and \ref{pcm0fig}, but only gave the result for the case when there was no turbulence. In this section we state the results for the case when turbulence is present, under the assumption that the turbulent fluctuations in the two branches are independent of each other (no relationship between $\alpha_1$ and $\alpha_2$).

Define a pair of new parameters, $\rho_i$, for $i=1,2$: \begin{equation}{1\over {\rho_i}}={1\over {r_0^2}}+\alpha_i+{{ik}\over{2d}}.\end{equation} Then, for the apparatus of fig. \ref{chengfig} (no PCM), the coincidence rate is of the generic form given in Eq. (\ref{coincresult0}) with correlation factor $m$, resolution width $R$, and field of view $W$ are given by:
\begin{eqnarray}
2m_0 &=& -{{a_{12}/\Delta_2-k^2/\Delta_1r_c^2d^2}\over {a_1/\Delta_2+k^2/\Delta_1d^2\rho_2}}\\
2R^2_0 &=& \left( {{a_1}\over {\Delta_2}}+{{k^2}\over {\Delta_1 d^2\rho_2}}\right)^{-1}\\
2W^2_0 &=&\left[ \left( {{a_2}\over {\Delta_2}}+{{k^2}\over {\Delta_1d^2\rho_1^\ast }}\right)
+\left( {{a_1}\over {\Delta_2}}+{{k^2}\over {\Delta_1d^2\rho_2}}\right)\left(
{{a_{12}/\Delta_2-k^2/\Delta_1r_c^2d^2}\over {a_1/\Delta_2+k^2/\Delta_1d^2\rho_2}} \right)^2\right]^{-1} ,
\end{eqnarray} where
\begin{eqnarray}\Delta_1 &=& \left[ {1\over {4a_0^4}}\left( 1+ {{4a_0^2}\over {r_c^2}}\right) +{{k^2}\over {d^2}}\right] +4\left( {{\alpha_1+\alpha_2}\over {r_0^2}}+\alpha_1\alpha_2
+{{ik}\over {2d}}(\alpha_1-\alpha_2)\right) \\
%\Delta_2 &=& a_1 |\bm x_1|^2 + a_2 |\bm x_2|^2 +a_{12} \bm x_1\cdot \bm x_2\\
\Delta_2 &=&4\left({1\over{\rho_2^\ast}}-{{4\alpha_2^2}\over{\Delta_1\rho_1^\ast}}\right)
\left({1\over{\rho_1}}-{{4\alpha_1^2}\over{\Delta_1\rho_2}}\right)-{1\over{r_c^4}}
\left(1+{{4\alpha_1\alpha_2}\over{\Delta_1}}\right)^2 \\
a_1 &=& {{k^2}\over {d^2}}\left[ -{{2\alpha_2}\over {\Delta_1 r_c^4}}\left( 1+{{4\alpha_1\alpha_2}\over {\Delta_1}}\right)\left( 1-{{4\alpha_1}\over {\Delta_1\rho_2^\ast}}\right) \right. \nonumber \\ & & \qquad  \left. +{{4\alpha_2^2}\over {\Delta_1^2r_c^4}}\left( {1\over {\rho_1}}-{{4\alpha_1^2}\over {\Delta_1\rho_2}}\right)+\left( {1\over {\rho_2^\ast}}-{{4\alpha_2^2}\over {\Delta_1\rho_1^\ast}}\right)\left( 1-{{4\alpha_1}\over {\Delta_1\rho_2}}\right)^2 \right] \\
a_2 &=& {{k^2}\over {d^2}}
\left[ -{{2\alpha_1}\over {\Delta_1 r_c^4}}\left( 1+{{4\alpha_1\alpha_2}\over {\Delta_1}}\right)\left( 1-{{4\alpha_2}\over {\Delta_1\rho_1^\ast}}\right)\right. \nonumber \\ & & \qquad \left. +{{4\alpha_1^2}\over {\Delta_1^2r_c^4}}\left( {1\over {\rho_2^\ast}}-{{4\alpha_2^2}\over {\Delta_1\rho_1^\ast}}\right)+\left( {1\over {\rho_1}}-{{4\alpha_1^2}\over {\Delta_1\rho_2}}\right)\left( 1-{{4\alpha_2}\over {\Delta_1\rho_1^\ast}}\right)^2 \right] \\
a_{12} &=& {{k^2}\over {d^2}}\left\{  -{1\over{r_c^2}}\left( 1+{{4\alpha_1\alpha_2}\over {\Delta_1}}\right)\left[ \left( 1-{{4\alpha_2}\over {\Delta_1 \rho_1^\ast}}\right)
\left( 1-{{4\alpha_1}\over {\Delta_1 \rho_2}}\right) +{{4\alpha_1\alpha_2}\over {\Delta_1^2r_c^4}}\right] \right. \\
&  &\quad \left. +{4\over {\Delta_1r_c^2}}\left[ \alpha_2\left( {1\over {\rho_1}}-{{4\alpha_1^2}\over{\Delta_1\rho_2}}\right)\left( 1-{{4\alpha_2}\over {\Delta_1 \rho_1^\ast}}\right)+\alpha_1\left( {1\over {\rho_2^\ast}}-{{4\alpha_2^2}\over{\Delta_1\rho_1 }}\right)\left( 1-{{4\alpha_1}\over {\Delta_1 \rho_2}}\right)\right]
\right\} \nonumber .
\end{eqnarray}

With the PCM added (fig. \ref{pcm0fig}), the corresponding expressions become:
\begin{eqnarray}
2m_{pcm} &=& -{{b_{12}/\Delta_4+k^2/\Delta_3r_c^2d^2}\over {b_1/\Delta_4+k^2/\Delta_3d^2\rho_2^\ast}}\\
2R^2_{pcm} &=& \left( {{b_1}\over {\Delta_4}}+{{k^2}\over {\Delta_3 d^2\rho_2^\ast}}\right)^{-1}\\
2W^2_{pcm} &=&\left[ \left( {{b_2}\over {\Delta_4}}+{{k^2}\over {\Delta_3d^2\rho_1^\ast }}\right)
-{1\over 4}\left( {{b_1}\over {\Delta_4}}+{{k^2}\over {\Delta_3d^2\rho_2^\ast}}\right)\left(
{{b_{12}/\Delta_4+k^2/\Delta_3r_c^2d^2}\over {b_1/\Delta_4+k^2/\Delta_3d^2\rho_2^\ast}} \right)^2\right]^{-1} ,
\end{eqnarray} with
\begin{eqnarray}\Delta_3 &=& \left[ {1\over {4a_0^4}}\left(1+ {{4a_0^2}\over {r_c^2}}\right) -{{k^2}\over {d^2}}-{{4ik}\over {dr_0^2}}\right] +4\left( {{\alpha_1+\alpha_2}\over {r_0^2}}+\alpha_1\alpha_2
-{{ik}\over {2d}}(\alpha_1+\alpha_2)\right) \\
%\Delta_4 &=& b_1 |\bm x_1|^2 + b_2 |\bm x_2|^2 +b_{12} \bm x_1\cdot \bm x_2\\
\Delta_4 &=&4\left({1\over{\rho_2}}-{{4\alpha_2^2}\over{\Delta_3\rho_1^\ast}}\right)
\left({1\over{\rho_1}}-{{4\alpha_1^2}\over{\Delta_3\rho_2^\ast}}\right)-{1\over{r_c^4}}
\left(1+{{4\alpha_1\alpha_2}\over{\Delta_3}}\right)^2 \\
b_1 &=& {{k^2}\over {d^2}}\left[ {{2\alpha_2}\over {\Delta_3 r_c^4}}\left( 1+{{4\alpha_1\alpha_2}\over {\Delta_3}}\right)\left( 1+{{4\alpha_1}\over {\Delta_3\rho_2^\ast}}\right) \right. \nonumber \\ & & \qquad  \left. +{{4\alpha_2^2}\over {\Delta_3^2r_c^4}}\left( {1\over {\rho_1}}-{{4\alpha_1^2}\over {\Delta_3\rho_2^\ast}}\right)+\left( {1\over {\rho_2}}-{{4\alpha_2^2}\over {\Delta_3\rho_1^\ast}}\right)\left( 1+{{4\alpha_1}\over {\Delta_3\rho_2^\ast}}\right)^2 \right] \\
b_2 &=& {{k^2}\over {d^2}}
\left[ {{2\alpha_1}\over {\Delta_3 r_c^4}}\left( 1+{{4\alpha_1\alpha_2}\over {\Delta_3}}\right)\left( 1+{{4\alpha_2}\over {\Delta_3\rho_1^\ast}}\right) \right. \nonumber \\ & & \qquad \left. +{{4\alpha_1^2}\over {\Delta_3^2r_c^4}}\left( {1\over {\rho_2}}-{{4\alpha_2^2}\over {\Delta_3\rho_1^\ast}}\right)+\left( {1\over {\rho_1}}-{{4\alpha_1^2}\over {\Delta_3\rho_2^\ast}}\right)\left( 1+{{4\alpha_2}\over {\Delta_3\rho_1^\ast}}\right)^2 \right] \\
b_{12} &=& {{k^2}\over {d^2}}\left\{  {1\over{r_c^2}}\left( 1+{{4\alpha_1\alpha_2}\over {\Delta_3}}\right)\left[ \left( 1+{{4\alpha_2}\over {\Delta_3 \rho_1^\ast}}\right)
\left( 1+{{4\alpha_1}\over {\Delta_3 \rho_2^\ast}}\right) +{{4\alpha_1\alpha_2}\over {\Delta_3^2r_c^4}}\right] \right. \\
&  &\quad \left. +{4\over {\Delta_3r_c^2}}\left[ \alpha_2\left( {1\over {\rho_1}}-{{4\alpha_1^2}\over{\Delta_3\rho_2^\ast }}\right)\left( 1+{{4\alpha_2}\over {\Delta_3 \rho_1^\ast}}\right)+\alpha_1\left( {1\over {\rho_2}}-{{4\alpha_2^2}\over{\Delta_3\rho_1^\ast }}\right)\left( 1+{{4\alpha_1}\over {\Delta_3 \rho_2^\ast}}\right)\right]
\right\} \nonumber .
\end{eqnarray}

In the absence of turbulence, $\alpha_1=\alpha_2=0$, we find that ${1\over{\rho_1}}={1\over{\rho_2}}={1\over{r_0^2}}+{{ik}\over {2d}}$, $\Delta_1=\Delta_2=\Delta_0$, and $\Delta_3=\Delta_4^\ast =\Delta_0-{{2k^2}\over {d^2}}+{{4ik}\over{r_0^2d}}$, where $\Delta_0$ is as defined in Eq. (\ref{delta0}). Consequently, it follows in a straightforward manner that the results stated in this appendix reduce back to the turbulence-free expressions given in section \ref{turbindep}.

\section*{Acknowledgments}

The authors would like to thank Dr. Jeffrey Shapiro for his very helpful comments.
This research was supported by the DARPA InPho program through the US
Army Research Office award W911NF-10-1-0404, and by a grant from Capella Intelligent
Systems, Inc.

\end{document}